\documentclass[twocolumn,
reprint,
groupedaddress,
amsmath,
amssymb,
aps,
prb,
]{revtex4}

\usepackage{graphicx}
\usepackage{dcolumn}
\usepackage{bm}

\begin{document}


\title{Entanglement spectrum and entangled modes\\ of random XX spin chains}

\author{Mohammad Pouranvari}
\email{pouranvari@magnet.fsu.edu}
\affiliation{National High Magnetic Field Laboratory and Department of Physics, Florida State University, Tallahassee, Florida 32306, USA}

\author{Kun Yang}
\email{kunyang@magnet.fsu.edu}
\affiliation{National High Magnetic Field Laboratory and Department of Physics, Florida State University, Tallahassee, Florida 32306, USA}

\date{\today}

\newcommand{\ket}[1]{\left| #1 \right>} 
\newcommand{\bra}[1]{\left< #1 \right|} 
\newcommand{\braket}[2]{\left< #1 \vphantom{#2} \right| \left. #2 \vphantom{#1} \right>} 

\begin{abstract}
We study in this work the ground state entanglement properties of finite XX spin-1/2 chains with random couplings, using
Jordan-Wigner transformation. We divide the system into two parts and study reduced density matrices (RDMs) of its subsystems. Due to the
free-fermion nature of the problem, the RDMs take the form of that of a free fermion thermal ensemble. Finding spectrum of the corresponding
entanglement Hamiltonian and corresponding eigenvectors, and comparing them with real space renormalization group (RSRG) treatment, we
establish the validity of the RSRG approach for entanglement in the limit of strong disorder, but also find its limitations when disorder is
weak. In the latter case our work provides a way to visualize the ``effective spins" that form long distance singlet pairs.
\end{abstract}

\maketitle


\section{Introduction}{\label{intro}}

Presence of entanglement is perhaps the most fundamental difference between quantum and classical physics. Recently entanglement has been widely used to characterize phases and phase transitions in condensed matter physics. Among various ways to quantify entanglement, the most frequently used is the entanglement entropy between two subsystems, obtained in the following manner.
For a quantum system in a pure state $\ket{\psi}$, the density matrix $\rho=\ket{\psi} \bra{\psi}$ contains all information of the system in
that state. Dividing the system into two subsystems $A$ and $B$ (most often in real space, which is the focus of this work), we can obtain the reduced density matrix (RDM) of each subsystem by tracing over degrees of freedom of the other subsystem: $\rho^{A/B}=tr_{B/A} (\rho)$. The entanglement entropy is $S_{E}^{A}=-tr(\rho^{A} \ln{\rho^{A}})$ (and we can obtain it for subsystem $B$ too, which for a pure state $ S_{E}^{A}=S_{E}^{B}$). While this \emph{number}
indeed describes how much subsystem $A$ and subsystem $B$ are entangled, it was pointed out by Li and Haldane \cite{lihaldane} that much more information about the system, especially the nature of its phase, is in the spectrum of the reduced density matrices $\rho^{A/B}$.

It seems natural to ask if even more useful information is available in the eigenstates of $\rho^{A/B}$. The answer is obviously yes, as they are the basis of numerical implementation of density matrix renormalization group. Here we ask this question from a different perspective, namely we would like to ask if one can extract some physical properties of the system from these eigenstates. We will show, using a very simple example, that the answer is still yes. The model is an antiferromagnetic spin-1/2 chain with random nearest neighbor XX couplings, which can be mapped onto a free fermion model with random hopping. Due to the Slater-determinant nature of the ground state,
\begin{equation} \label{rho}
\rho^{A}=\frac{1}{Z} e^{-H^{A}}
\end{equation}
is characterized by a free-fermion {\em entanglement} Hamiltonian $H^A$. It is easy to obtain (at least numerically) the {\em single particle} eigenmodes (which make up, but are different from the multi-particle eigenstates of $\rho^{A}$), and their counterparts in part $B$; we call such pairs entangled (single-particle) modes. For the present model these entangled modes represent singlet pairs formed by effective spins, often over long distance (see below).

This model is a special case of random XXZ model, which could also be studied using the real space renormalization group (RSRG, also referred to as strong disorder RG in literature).\cite{dsfisher} Within this {\em approximation}, it was found that the long-distance, low-energy behavior of this entire class of models is dominated by the random singlet fixed point, in which singlet bonds form between spins in opposite sub-lattices over arbitrarily large distances. Refael and Moore\cite{refael} calculated the entanglement entropy of the random singlet phase using the RSRG method. Within this approximation scheme, entanglement between $A$ and $B$ comes from singlet pairs formed by a spin in $A$ and another spin in $B$. If it were exactly correct, then in the fermion language above there should be a zero mode associated with each of such pair, and the mode wave function is (ultra)localized at the sites of these entangled spins. In our exact diagonalization study of the random XX model we find, of course, this is {\em not} the case unless the randomness is very strong (which is the regime RSRG works well); in general the (entanglement) ``energies" of these modes are not zero, and the mode wave functions do have finite extent. We do find that for modest disorder strength, the entanglement ``energies" of the low-lying entangled modes (corresponding to long bonds that cross the subsystem boundary) indeed approaches zero with increasing system size and bond length, supporting the asymptotically free nature of RSRG (including in entanglement calculation). On the other hand the spatial extent of these modes do {\em not} decrease with such increase in bond length. These mode wave functions are understood as real space images of the {\em effective} spins that form singlets over long distance; the profiles of such effective spins cannot be obtained within RSRG. Thus entanglement provides a new way to probe such effective spins.

In a recent work, Fagotti, et al \cite{fagotti} also used the free fermion mapping to calculate disorder-averaged moments of entanglement spectrum in this model. A number of other papers\cite{Laflorencie,Santachiara,hoyos,tran,RefaelMooreReview} studied entanglement entropy of random Heisenberg and other closely related models using various numerical methods. Our work is complementary to Ref. \onlinecite{fagotti} and the other earlier works as our emphasis is on entangled {\em modes} of specific realizations of certain random distributions of couplings, not just disorder-averaged quantities.

The remainder of the paper is organized as the following. In Sec. II we introduce our model and notions like entanglement energy and entangled modes. Sec. III is a brief review of RSRG from the viewpoint of entanglement. Our numerical results are presented and analyzed in Sec. IV. Sec. V offers a summary and some concluding remarks.

\section{Model and Entangled Modes} {\label{mainbody}}

The model we work with is a one-dimensional (1D) spin-1/2 antiferromagnetic XX model with $N$ sites and with random nearest neighbour couplings. The Hamiltonian of the system is

\begin{equation} \label{origham}
H=\sum_{n=1} ^{N-1} J_{n} (s_{n}^{x} s_{n+1}^{x}+s_{n}^{y}s_{n+1}^{y}) .
\end{equation}
Note that we use open instead of periodic boundary condition here.\cite{pbcnote} By mapping spin operators to fermion operators via Jordan-Wigner transformation:
\begin{equation}
c_{n}=e^{(i \pi \sum_{j<n} s_{j}^{+}s_{j}^{-})} s_{n}^{-} ,
\end{equation}
where $c$ is a fermionic operator, the Hamiltonian becomes

\begin{equation} \label{hamc}
H=\frac{1}{2} \sum_{n=1} ^{N-1} J_{n} (c_{n}^{\dagger} c_{n+1}+c_{n+1}^{\dagger}c_{n}).
\end{equation}
The $(N-1)$ $J$'s in Eq. (\ref{hamc}) will be generated based on random distribution functions to be specified later.
It is a special case of the most general free Fermi Hamiltonian:
\begin{equation} \label{hamh}
H=\sum _{i=1}^{N}\sum _{j=1}^{N} h_{ij} c_{i}^{\dagger} c_{j} ,
\end{equation}
with
\begin{equation}
h_{ij} =
\left\{
	\begin{array}{ll}
		(1/2)J_{i}  & \mbox{if } j=i+1 \\
		(1/2)J_{i-1} & \mbox{if } j=i-1
	\end{array}
\right.
\end{equation}
and $h_{ij} =0$ otherwise.

The number of fermions is related to magnetization:
\begin{equation}
N_{F}=\frac{1}{2} N + \sum_{n=1}^{N} s_{n} ^{z};
\end{equation}
for the ground state we expect $N_F=N/2$ (we always work with even $N$).

Now, we divide the system into two subsystems $A$ and $B$, often (but not necessarily) with equal number of sites. Subsystem $A$ is from site $1$ to $N_A$, and subsystem $B$ is from site $N_A+1$ to $N$. We know that we can write the RDM $\rho^A$ (since it is a positive definite operator) as the exponential of a Hermitian operator, $H^A$, as in Eq. (\ref{rho}). For the special case of free fermion system here, the entanglement Hamiltonian $H^A$ is also a {\em free fermion} Hamiltonian:\cite{rdm}

\begin{equation} \label{HA}
H^{A}=\sum _{i,j=1}^{N_A} h_{ij}^A c_{i}^{\dagger} c_{j}.
\end{equation}

To determine eigenmodes and eigenvalues of $h^{A}$ we follow Ref. \onlinecite{correl} by defining correlation function

\begin{equation}  \label{cor}
C_{ij}^A=<c_{i}^{\dagger} c_{j}>,
\end{equation}
which is an $N_A\times N_A$ symmetric matrix. Its $N_A$ eigenmodes are the same as those of $h^A$, while the corresponding eigenvalues are related to each other:
\begin{equation} \label{zeta}
n_k^A=\frac{1}{1+e^{\epsilon_k ^A}},
\end{equation}
where ${\epsilon_k ^A}$ is an eigenvalue of $h^A$ referred to as entanglement energy (not to be confused with eigenvalues of the original Hamiltonian $h$!), and $n_k$ is the probability of finding a fermion in the corresponding mode.

We may have three different types of $\epsilon$. First, it can be a very big negative number, which means the probability of finding a fermion at mode $k$ is almost 1; thus this fermion is almost exclusively localized in $A$. Second, it can be a very
big positive number, which means the probability of finding a fermion at mode $k$ is almost zero; this means that there is a fermion that is almost localized completely in the complementary mode (to be discussed later) localized in subsystem $B$. These modes contribute very little to entanglement. Third, it can be neither of the above (and possibly close to zero), which means there is a substantial probability $n^A$ less than 1 that this particle is in subsystem $A$. This {\em necessarily} implies that there is a corresponding mode in $B$ where the {\em same} particle resides with probability $n^B=1-n^A$. These pairs of modes, referred to as entangled modes, dominate entanglement. To find the relation between eigenvalues of these modes, we consider the case of a single fermion occupying a mode that is split into two parts. Since the total probability is
\begin{equation}\label{n}
n=n^A+n^B=\frac{1}{1+e^{\epsilon^A}}+\frac{1}{1+e^{\epsilon^B}}=1,
\end{equation}
we have $\epsilon^A=-\epsilon^B$. It is straightforward to establish the corresponding relation (\ref{n}) in the multi particle case: it follows from the Schmidt decomposition that $\rho^A$ and $\rho^B$, and thus $H^A$ and $H^B$ must have the same spectra, while the corresponding eigenstates must have their particle numbers add up to $N$; as a result the spectra of $h^A$ and $h^B$ are related by particle-hole transformation. In the following we will use these relations to identify such pairs of entangled modes.

In fact the entangled modes can be glued together to form a set of $N_F$ orthonormal modes occupied by the fermions in the ground state, and the weight of these modes in subsystems $A$ and $B$ correspond to $n^A$ and $n^B$ respectively. One way to obtain these modes directly (without using the RDMs) is using Klich's method\cite{klich} as outlined below.

We divide the Hilbert space of one particle states into two parts corresponding to our subsystem $A$ and $B$, and introduce corresponding projection operators $P_A$ and $P_B$ as:

\begin{equation}
P_A=\sum_{i=1}^{N_A} \ket{i} \bra{i};\hskip 0.5 cm P_B=\sum_{i=N_A+1}^{N} \ket{i} \bra{i}.
\end{equation}
Introduce a Hermitian matrix
\begin{equation}
M_{kk'}=\braket{P_A k'}{P_A k},
\end{equation}
where $\ket{k}$'s are the lowest $N_F$ eigenvectors of matrix $h$ in Eq. (\ref{hamh}) {\em occupied} by fermions in the ground state. Then diagonalize this $N_F\times N_F$ matrix by a unitary matrix $V$ as $M = V d {V}^{\dagger}$, we obtain these new modes:
\begin{eqnarray}
\ket{l}^A&=&\frac{1}{\sqrt{d_l}} \sum_k V^{\dagger}_{lk} P_A \ket{k} \nonumber \\
               &=&\frac{1}{\sqrt{d_l}} \sum_{k=1}^{N_F} \sum_{i=1}^{N_A} V^{\dagger}_{lk} U_{ik} \ket{i}, \nonumber \\
\end{eqnarray}

\begin{eqnarray}
\ket{l}^B&=&\frac{1}{\sqrt{1-d_l}} \sum_k V^{\dagger}_{lk} P_B \ket{k} \nonumber \\
               &=&\frac{1}{\sqrt{1-d_l}} \sum_{k=1}^{N_F} \sum_{i=N_A+1}^{N} V^{\dagger}_{lk} U_{ik} \ket{i}, \nonumber \\
\end{eqnarray}
where $U$ is the unitary matrix that diagonalize matrix $h$ in Eq. (\ref{hamh}) and $l$ goes from $1$ to $N_F$. $\ket{l}^A$ corresponds to eigenvalue $d_l$ (which is the same as $n_l^A$  when $N_F=N_A$) and it is in subsystem $A$. $\ket{l}^B$ corresponds to eigenvalue $1- d_l$ and it is in subsystem $B$. Now, by sticking these modes together we obtain a mode in whole system (Klich eigenmodes):
\begin{equation}
\ket{l}=\sqrt{d_l} \ket{l}^A +\sqrt{1-d_l} \ket{l}^B=\sum_{k=1}^{N_F} V^{\dagger}_{lk} \ket{k}.
\end{equation}
As shown by Klich,\cite{klich} the single Slater determinant formed by these modes is identical to the original ground state, while projecting them to subsystems A and B gives rise to the entangled modes with proper weight. These Klich eigenmodes are thus particularly useful in studies of entanglement.

We note in general the Klich eigenmodes are {\em not} the same as the occupied eigenmodes of Hamiltonian matrix $h$ in Eq. (\ref{hamh}). We illustrate this by calculating the $N_F \times N_F$ overlap matrix between these modes, in which each row represents an eigenmode of matrix $h$ and each column represents one particle Klich eigenmode. Table \ref{sdis} is the overlap matrix of a sample with $N=30, N_F=15, \alpha=0.1$ (see below for definition of $\alpha$ and its physical meaning). In this case of strong disorder, each Klich eigenmode has large overlap(s) with one (or at most two) eigenmode(s) of matrix $h$, and one can (almost) identify a one-to-one correspondence between them. Table \ref{wdis} is overlap matrix of a sample with $N=30, N_F=15, \alpha=0.9$. In this case of weaker disorder, such a correspondence is not possible.

\begin{table*}
\begin{ruledtabular}
\caption{\label{sdis} Overlaps between Klich eigenmodes and filled eigenmodes of original Hamiltonian ($N_F \times N_F$ matrix) when $N=30, N_F=15, \alpha=0.1$. Each row represents an eigenmode of matrix $h$ and each column represents one particle Klich eigenmode. One can find a one-to-one correspondence between them in most cases. }
\begin{tabular}{cccccccccccccccccccc}
$0.00$ & $0.00$ & $0.00$ & $0.00$ & $0.00$ & $0.00$ & $0.00$ & $0.00$ & $0.00$ & $0.00$ & $0.00$ & $0.00$ & $0.00$ & $0.00$ & $\textbf{1.00}$ \\
$0.00$ & $0.00$ & $0.00$ & $0.00$ & $0.00$ & $0.00$ & $0.00$ & $0.00$ & $0.00$ & $0.00$ & $0.00$ & $\textbf{0.94}$ & $0.05$ & $0.00$ & $0.00$ \\
$0.00$ & $0.00$ & $0.00$ & $0.00$ & $0.00$ & $0.00$ & $0.00$ & $0.00$ & $0.00$ & $0.00$ & $0.00$ & $0.00$ & $0.00$ & $\textbf{1.00}$ & $0.00$ \\
$0.00$ & $0.00$ & $0.00$ & $0.00$ & $0.00$ & $0.02$ & $\textbf{0.79}$ & $0.00$ & $\textbf{0.19}$ & $0.00$ & $0.00$ & $0.00$ & $0.00$ & $0.00$ & $0.00$ \\
$\textbf{1.00}$ & $0.00$ & $0.00$ & $0.00$ & $0.00$ & $0.00$ & $0.00$ & $0.00$ & $0.00$ & $0.00$ & $0.00$ & $0.00$ & $0.00$ & $0.00$ & $0.00$ \\
$0.00$ & $\textbf{1.00}$ & $0.00$ & $0.00$ & $0.00$ & $0.00$ & $0.00$ & $0.00$ & $0.00$ & $0.00$ & $0.00$ & $0.00$ & $0.00$ & $0.00$ & $0.00$ \\
$0.00$ & $0.00$ & $\textbf{0.81}$& $\textbf{0.19}$ & $0.00$ & $0.00$ & $0.00$ & $0.00$ & $0.00$ & $0.00$ & $0.00$ & $0.00$ & $0.00$ & $0.00$ & $0.00$ \\
$0.00$ & $0.00$ & $0.00$ & $0.00$ & $0.00$ & $0.00$ & $0.00$ & $0.00$ & $0.00$ & $0.05$ & $\textbf{0.95}$ & $0.00$ & $0.00$ & $0.00$ & $0.00$ \\
$0.00$ & $0.00$ & $0.00$ & $0.00$ & $0.00$ & $0.02$ & $\textbf{0.17}$ & $0.00$ & $\textbf{0.80}$ & $0.00$ & $0.00$ & $0.00$ & $0.00$ & $0.00$ & $0.00$ \\
$0.00$ & $0.00$ & $0.00$ & $0.00$ & $0.00$ & $0.00$ & $0.00$ & $0.00$ & $0.00$ & $0.00$ & $0.00$ & $0.06$ & $\textbf{0.94}$ & $0.00$ & $0.00$ \\
$0.00$ & $0.00$ & $\textbf{0.19}$ & $\textbf{0.81}$ & $0.00$ & $0.00$ & $0.00$ & $0.00$ & $0.00$ & $0.00$ & $0.00$ & $0.00$ & $0.00$ & $0.00$ & $0.00$ \\
$0.00$ & $0.00$ & $0.00$ & $0.00$ & $0.00$ & $\textbf{0.96}$ & $0.04$ & $0.00$ & $0.01$ & $0.00$ & $0.00$ & $0.00$ & $0.00$ & $0.00$ & $0.00$ \\
$0.00$ & $0.00$ & $0.00$ & $0.00$ & $0.00$ & $0.00$ & $0.00$ & $0.00$ & $0.00$ & $\textbf{0.95}$ & $0.05$ & $0.00$ & $0.00$ & $0.00$ & $0.00$ \\
$0.00$ & $0.00$ & $0.00$ & $0.00$ & $\textbf{1.00}$ & $0.00$ & $0.00$ & $0.00$ & $0.00$ & $0.00$ & $0.00$ & $0.00$ & $0.00$ & $0.00$ & $0.00$ \\
$0.00$ & $0.00$ & $0.00$ & $0.00$ & $0.00$ & $0.00$ & $0.00$ & $\textbf{1.00}$ & $0.00$ & $0.00$ & $0.00$ & $0.00$ & $0.00$ & $0.00$ & $0.00$
\end{tabular}
\end{ruledtabular}
\end{table*}

\begin{table*}
\begin{ruledtabular}
\caption{\label{wdis} Overlaps between Klich eigenmodes and filled eigenmodes of original Hamiltonian ($N_F \times N_F$ matrix) when $N=30, N_F=15, \alpha=0.9$. Each row represents an eigenmode of matrix $h$ and each column represents one particle Klich eigenmode. The overlaps spread among quite a few eigenmodes, and a one-to-one correspondence between them is not possible.}
\begin{tabular}{cccccccccccccccccccc}
$0.00$ & $0.00$ & $0.00$ & $0.00$ & $0.00$ & $0.02$ & $0.04$ & $0.00$ & $0.05$ & $\textbf{0.15}$ & $\textbf{0.26}$ & $\textbf{0.42}$ & $0.04$ & $0.00$ & $0.00$ \\
$0.00$ & $0.00$ & $0.00$ & $0.00$ & $0.00$ & $0.00$ & $0.00$ & $0.00$ & $0.00$ & $0.00$ & $0.00$ & $0.00$ & $0.00$ & $\textbf{0.99}$ & $0.00$ \\
$\approx 0$ & $0.00$ & $0.00$ & $0.00$ & $0.00$ & $\textbf{0.27}$ & $\textbf{0.35}$ & $0.01$ & $\textbf{0.18}$ & $0.01$ & $0.01$ & $\textbf{0.15}$ & $0.03$ & $0.00$ & $0.00$ \\
$\textbf{0.28}$ & $\textbf{0.71}$ & $0.00$ & $0.00$ & $0.00$ & $0.00$ & $0.00$ & $0.00$ & $0.00$ & $0.00$ & $0.00$ & $0.00$ & $0.00$ & $0.00$ & $0.00$ \\
$\approx 0$ & $0.00$ & $0.00$ & $0.00$ & $0.00$ & $0.00$ & $0.00$ & $0.00$ & $0.00$ & $0.00$ & $0.00$ & $0.00$ & $0.00$ & $0.00$ & $\textbf{1.00}$ \\
$\approx 0$ & $0.00$ & $\textbf{0.80}$ & $\textbf{0.19}$ & $0.01$ & $0.00$ & $0.00$ & $0.00$ & $0.00$ & $0.00$ & $0.00$ & $0.00$ & $0.00$ & $0.00$ & $0.00$ \\
$\textbf{0.71}$ & $\textbf{0.29}$ & $0.00$ & $0.00$ & $0.00$ & $0.00$ & $0.00$ & $0.00$ & $0.00$ & $0.00$ & $0.00$ & $0.00$ & $0.00$ & $0.00$ & $0.00$ \\
$\approx 0$ & $0.00$ & $0.00$ & $0.00$ & $0.00$ & $0.08$ & $0.03$ & $0.00$ & $0.00$ & $\textbf{0.20}$ & $\textbf{0.22}$ & $\textbf{0.12}$ & $\textbf{0.35}$ & $0.00$ & $0.00$ \\
$\approx 0$ & $0.00$ & $\textbf{0.20}$ & $\textbf{0.74}$ & $0.05$ & $0.00$ & $0.00$ & $0.00$ & $0.00$ & $0.00$ & $0.00$ & $0.00$ & $0.00$ & $0.00$ & $0.00$ \\
$\approx 0$ & $0.00$ & $0.00$ & $0.01$ & $0.00$ & $0.05$ & $0.00$ & $0.00$ & $0.02$ & $0.08$ & $0.02$ & $\textbf{0.27}$ & $\textbf{0.56}$ & $0.00$ & $0.00$ \\
$0.00$ & $0.00$ & $0.00$ & $0.00$ & $0.00$ & $\textbf{0.44}$ & $0.04$ & $0.01$ & $\textbf{0.34}$ & $0.00$ & $\textbf{0.12}$ & $0.03$ & $0.01$ & $0.00$ & $0.00$ \\
$0.00$ & $0.00$ & $0.00$ & $0.00$ & $0.06$ & $0.08$ & $\textbf{0.25}$ & $0.00$ & $0.04$ & $\textbf{0.30}$ & $\textbf{0.25}$ & $0.01$ & $0.00$ & $0.00$ & $0.00$ \\
$0.00$ & $0.00$ & $0.00$ & $0.05$ & $\textbf{0.88}$ & $0.02$ & $0.01$ & $0.00$ & $0.00$ & $0.03$ & $0.02$ & $0.00$ & $0.00$ & $0.00$ & $0.00$ \\
$0.00$ & $0.00$ & $0.00$ & $0.00$ & $0.00$ & $0.05$ & $\textbf{0.27}$ & $0.01$ & $\textbf{0.33}$ & $\textbf{0.25}$ & $\textbf{0.10}$ & $0.00$ & $0.00$ & $0.00$ & $0.00$ \\
$0.00$ & $0.00$ & $0.00$ & $0.00$ & $0.00$ & $0.00$ & $0.00$ & $\textbf{0.97}$ & $0.03$ & $0.00$ & $0.00$ & $0.00$ & $0.00$ & $0.00$ & $0.00$
\end{tabular}
\end{ruledtabular}
\end{table*}

\section{Real Space Renormalization Group Method} {\label{rsrgmethod}}

\begin{figure}
\includegraphics[scale=.55]{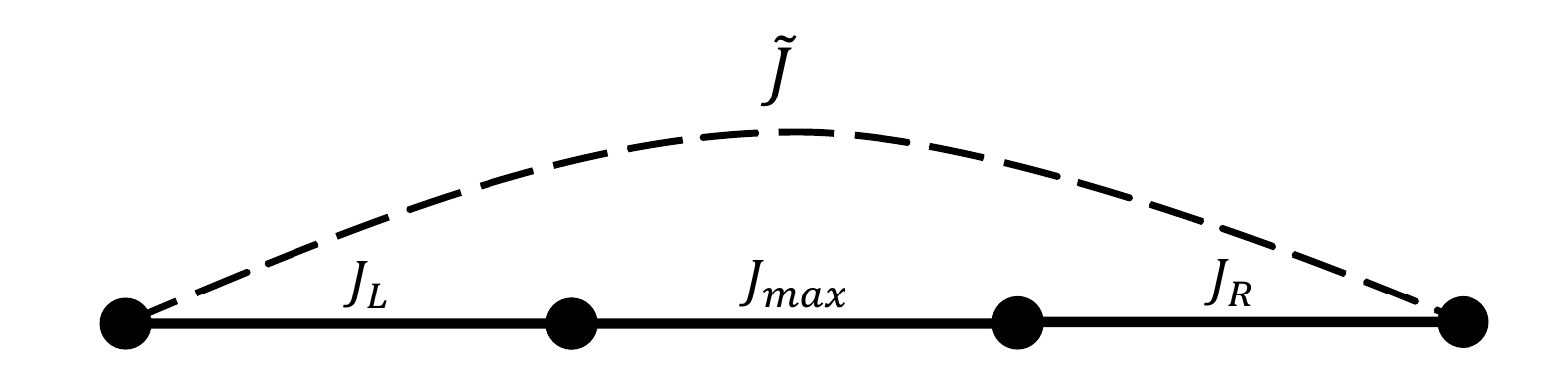}
\caption{\label{rsrg} A schematic plot real space renormalization group (RSRG) procedure. Each solid line with coupling $J$ connects two spins which are depicted by filled circles. The dashed line is an effective coupling generated by RSRG.}
\end{figure}

\begin{figure}
\includegraphics[width=.5\textwidth]{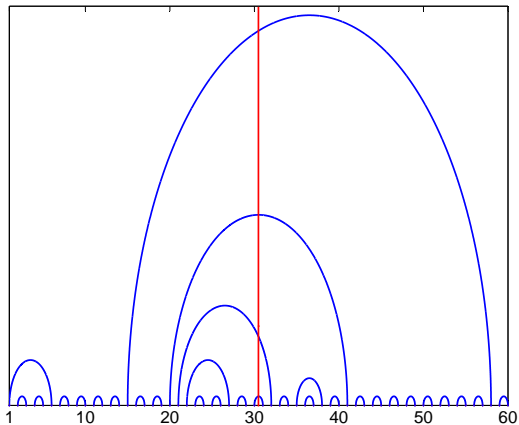}
\caption{\label{rsrg2}[Color online] An example of random singlet ground state formed in a spin chain of $N=60$, with blue curves connecting spins forming singlet pairs. Only those singlets that cross the boundary (red vertical line) contribute to entanglement.}
\end{figure}

We can obtain an {\em approximate} ground state of the Hamiltonian of Eq. (\ref{origham}) (with $J$'s distributed with a distribution function) by real space renormalization group (RSRG) method. As described by  Dasguptaand Ma \cite{dasgupta} and Fisher,\cite{dsfisher} for the Hamiltonian of Eq. (\ref{origham}), we first pick up the maximum $J=J_{max}$. Two spins that are coupled by this $J_{max}$ will form a singlet in the ground state of the Hamiltonian if we ignore their couplings to other spins. Then we remove the two spins that are coupled by this coupling constant. The two other couplings $J_L, J_R$ that couple this singlet pair with the rest of chain are treated perturbatively, resulting in a new coupling constant $\tilde{J}\approx\frac{J_L J_R}{J_{max}}$ between the further neighbor spins (see FIG.\ref{rsrg}). We note that this new coupling is smaller than $J_L$ and $J_R$, resulting in a reduction of overall energy scale. As we repeat this procedure many times we have singlet bonds that form between spins that are very far apart (thus forming a random singlet state). In the meantime the distribution of $J$'s broadens (with probability of finding smaller effective bonds increases), and approaches a power-law distribution function:

\begin{equation}\label{prob}
P(J) =
\left\{
	\begin{array}{ll}
		\alpha \ J^{-1+\alpha}  & \mbox{, if } 0  \leq J \leq 1 \\
		0                                   & \mbox{, otherwise}
	\end{array}
\right.
\end{equation}

with $\alpha$ slowly decreasing with energy scale. We thus use this type of distributions of $J$'s for our Hamiltonian in Eq. (\ref{origham}) or Eq. (\ref{hamc}), with smaller $\alpha$ corresponding to broader distribution (on logarithmic scale) and stronger disorder.  Fig. \ref{rsrg2} shows an example of the random singlet state generated by the RSRG procedure.

Such {\em approximate} ground state is made of a direct product of singlet pairs, and only those singlets that cross the boundary contribute to entanglement. In the fermion language introduced earlier, there is one fermion living on the two sites that form a singlet, with equal probability. Thus the entanglement energy of the eigenmodes of $h^A$ for subsystems $A$ would only take 3 possible values for such a state: $+\infty$, $-\infty$ and $0$; the last corresponds to singlets formed by spins on opposite sides of the boundary, with entangled modes having equal weight on the two sites of these spins, and {\em zero} weight elsewhere.  We will show that this is {\em not} exactly the case for the {\em exact} ground state.

\section{Results}{\label{results}}

As mentioned before, two important predictions of RSRG method are as follows. First, entangled modes are ultra localized at entangled spin sites; second, entanglement energy of these modes are {\em exactly zero}, and that corresponds to $n^A=n^B=0.5$. In this section we show that these predictions are not accurate; on the other hand the entanglement energy does approach zero asymptotically in the limit of large separation between the entangled modes, reflect the asymptotic exactness of the RSRG method.

\subsection{Entangled modes}{\label{em}}
Particle-hole symmetry of  Hamiltonian Eq. (\ref{origham}) forces the entanglement energies to be in $\pm$ pairs. \cite{cheonghenley} If we choose number of spin sites as: $N=2\times$ odd integer, then $N_A=N/2$ is odd,  we will have at least one pair of entangled modes whose entanglement energy is exactly zero. In this subsection we focus on these modes.

\begin{figure}
\includegraphics[width=.5\textwidth]{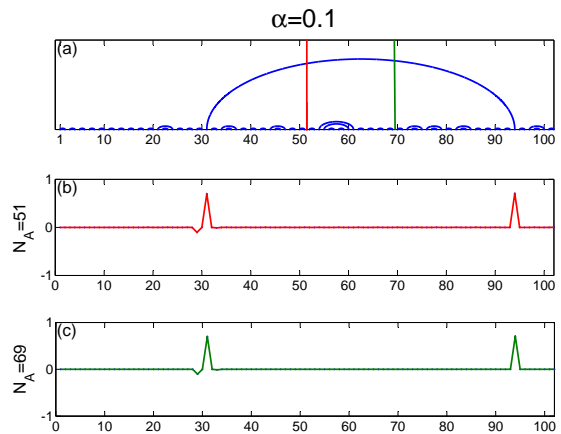}
\caption{\label{bound_102_p1}[Color online] (a) Singlet bond configuration generated by RSRG method for an $N=102$ sample with $\alpha=0.1$ (strong disorder). Panels (b) and (c) are zero energy Klich eigenmodes corresponding to two different boundaries. Panel (b): Zero (entanglement) energy Klich eigenmode which is strongly localized at the entangled spins (consistent with RSRG prediction), with $N_A=51$ (red line), and panel (c) corresponds to $N_A=69$ (green line).}
\end{figure}

Fig. \ref{bound_102_p1} shows a sample with $N=102$, and we choose $N_F=N_A=N_B=N/2=51$. $J$'s are generated using the distribution function of Eq. (\ref{prob}) with $\alpha=0.1$ (strong disorder). Panel (a) shows the singlet pair formation according to RSRG procedure. We can see there is just one singlet pair that crosses the boundary, and contributes to entanglement. On the other hand, we can also calculate eigenmodes and eigenvalues of entanglement Hamiltonian exactly. Panel (b) shows the zero (entanglement) energy mode, the Klich eigenmode with $n^A=n^B=0.5$. We see that this Klich eigenmode is indeed strongly localized around the two spins that are entangled (consistent with RSRG).

\begin{figure}
\includegraphics[width=.5\textwidth]{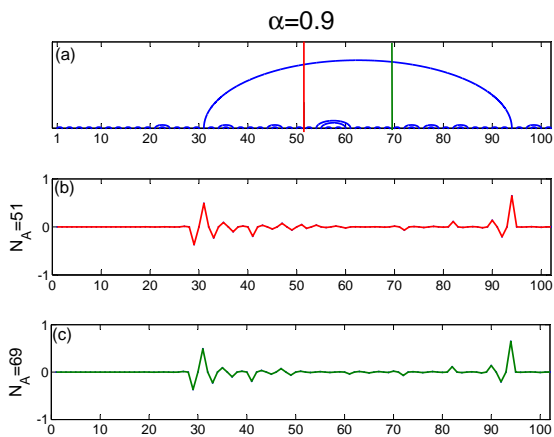}
\caption{\label{bound_102_p9}[Color online] (a) Singlet bond configuration generated by RSRG method for a sample with $\alpha=0.9$ (weaker disorder). Panels (b) and (c) are zero energy Klich eigenmodes corresponding to two different boundaries. Panel (b): Zero (entanglement) energy Klich eigenmode which is peaked at the entangled spins predicted by RSRG but much more spread out, $N_A=51$ (red line) and panel (c) corresponds to $N_A=69$ (green line).}
\end{figure}

For weaker disorder RSRG method may not be as reliable. Fig. \ref{bound_102_p9} shows a similar sample of Fig. \ref{bound_102_p1}, but with $\alpha=0.9$. The RSRG method generates almost the same singlet pair configuration [see panel (a)]. However in panel (b) we see the zero (entanglement) energy Klich eigenmode is much more spread out compared to the previous case, even though it still peaks at the location of the two spins RSRG predicts to form the entangled pair. In such cases (and including the earlier strong disorder example) we should view the entanglement as coming from two {\em effective} spins that actually spread over some finite spatial extent, whose profile are described by these entangled modes. This is appropriate as the profiles of these entangled modes are {\em not} sensitive to the position of the boundaries of these systems (see panel c of Figs. \ref{bound_102_p1} and \ref{bound_102_p9}), indicating that they are intrinsic properties of the degrees of freedom that are entangled over long distance. We note that such information about the effective spin(s) (that form singlets) cannot be obtained from RSRG, but it may also be accessed by looking at spin-spin correlation functions in an exact numerical calculation.\cite{hoyos}

To illustrate the point above quantitatively, we consider the same samples of $N=102$, but choose different locations of the boundary: $N_A=19, 39, 51, 69, 79, 89$, and calculate the overlaps of zero energy Klich eigenmodes among them, forming a $6 \times 6$ overlap matrix. The results are presented in Table \ref{t_p1_overlap} for $\alpha=0.1$ and in Table \ref{t_p9_overlap} for $\alpha=0.9$. RSRG predicts the same entangled spin pair for them, except for the first choice of $N_A=19$. As a result the overlaps are all close to 1 (closer to 1 for the stronger disorder case $\alpha=0.1$), except for those involving $N_A=19$.

\begin{table*}
\caption{\label{t_p1_overlap} Overlaps between zero energy Klich eigenmodes  for a sample with $\alpha=0.1$, $N=102, N_F=51$ corresponding to different choices of boundary, $N_A=19, 39, 51, 69, 79, 89$. The first one, $N_A=19$, corresponds to a different entanglement configuration.}
\begin{tabular}{c|cccccc}
& $N_A=19$&$N_A=39$&$N_A=51$&$N_A=69$&$N_A=79$&$N_A=89$ \\
\hline
$N_A=19$&1.0                             & 0    &0 & 0& 0& 0 \\
$N_A=39$& 0&                  1.0               &     0.99                        &     0.99                       &   0.99                         &    0.99 \\
$N_A=51$& 0&     0.99                          &                 1.0              &    1.0                         &   0.99                         &    0.99 \\
$N_A=69$& 0&     0.99                          &   1.0                            &    1.0                         &   0.99                         &    0.99 \\
$N_A=79$& 0&    0.99                           &    0.99                         &    0.99                       &   1.0                           &    0.99 \\
$N_A=89$& 0&    0.99                           &    0.99                         &    0.99                       &   0.99                         &    1.0 \\
\end{tabular}
\end{table*}

\begin{table*}
\caption{\label{t_p9_overlap} Overlaps between zero energy Klich eigenmodes for a sample with $\alpha=0.9$, $N=102, N_F=51$ corresponding to different choices of boundary, $N_A=19, 39, 51, 69, 79, 89$. The first one, $N_A=19$, corresponds to a different entanglement configuration.}
\begin{tabular}{c|cccccc}
& $N_A=19$&$N_A=39$&$N_A=51$&$N_A=69$&$N_A=79$&$N_A=89$ \\
\hline
$N_A=19$&1.0&   $3.1\times 10^{-6}$ &  $3.2\times 10^{-6}$ &  $3.2\times 10^{-6}$&   $3.1\times 10^{-6}$ &  $3.0\times 10^{-6}$ \\
$N_A=39$&        $3.1\times 10^{-6}$ &            1.0&      0.841&       0.843&      0.834 &     0.812 \\
$N_A=51$&     $3.2\times 10^{-6}$  &    0.841&            1.0 &     0.993&      0.982 &     0.956 \\
$N_A=69$&     $3.2\times 10^{-6}$      & 0.843   &   0.993   &         1.0   &   0.988 &     0.962\\
$N_A=79$&    $3.1\times 10^{-6}$    &  0.834 &    0.982     & 0.988 &           1.0   &   0.968  \\
$N_A=89$&    $3.0\times 10^{-6}$    &  0.812  &    0.956 &     0.962  &    0.968  &          1.0 \\
\end{tabular}
\end{table*}

To quantify the spatial extent of such effective spins, we use inverse participation ratio (IPR) of the corresponding Klich eigenmode:

\begin{equation}
\text{IPR}=\frac{1}{\sum_i | \psi_i |^4}.
\end{equation}
For example, if a wave function is localized over just one site, $\psi_{i}=\delta_{ij}$, then IPR$=1$, and if a wave function spreads equally over all sits, $\psi_{i}=1/\sqrt{N}$, then IPR$=N$; thus IPR is a measure of how much the wave function spreads. In particular, if RSRG were exact, it would predict IPR=$2$ for the Klich modes that correspond to the singlet pair contributing to entanglement. It is thus reasonable to identify $\frac{IPR}{2}$ as the spatial extent of an effective spin that contribute to entanglement, through formation of a singlet with another effective spin in the other subsystem.

\begin{figure}
\includegraphics[width=.5\textwidth]{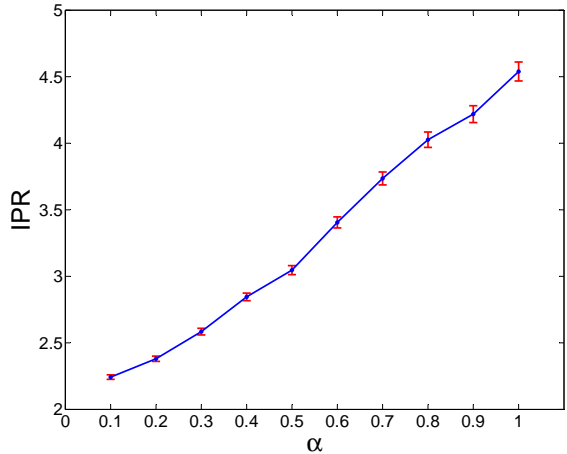}
\caption{\label{IPR_alpha_82_1000}[Color online] Inverse participation ratio (IPR) of the zero (entanglement) energy Klich modes versus $\alpha$, averaged over 1000 samples. The chain length is fixed to be 82. Standard error is the red line.}
\end{figure}

To see how IPR depends on $\alpha$, we fix the chain length to be $N=2\times 41=82, N_A=N_B=N_F=41$, and vary  $\alpha$ between $0.1$ and $1$. We only keep samples in which we have only one singlet bond crossing the boundary (and contributing to entanglement) according to RSRG. Result is shown in Fig. \ref{IPR_alpha_82_1000}.  We see IPR is very close to $2$ for strong disorder, indicating the entangled modes are strongly localized on {\em individual} spins (as RSRG predicts). On the other hand it is also clear that IPR increases with increasing $\alpha$ (or decreasing disorder strength), indicating the entangled modes are spreading out over multiple spins, a piece of physics RSRG is unable to capture.

\begin{figure}
\includegraphics[width=.5\textwidth]{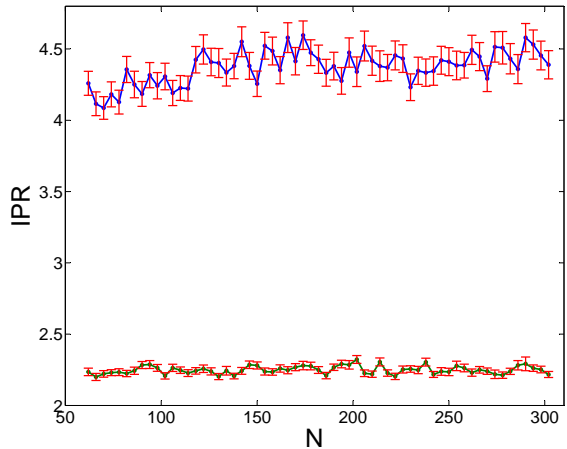}
\caption{\label{IPR_N_62_302_4_500}[Color online] IPR of zero entanglement energy Klich modes averaged over $500$ samples, for two values of $\alpha$, versus number of sites $N$. Green line is average IPR for strong disorder, $\alpha=0.1$, and blue line is average IPR for weak disorder, $\alpha=0.9$. Standard error is the red line.}
\end{figure}

We find that the spatial spreading of the entangled modes is determined by the disorder strength, and insensitive to system size or length of the entangled bond (which typically grows with system size). To demonstrate this point we calculate the average IPR of the zero (entanglement) energy Klich mode versus number of spins, $N$, for fixed $\alpha$. Again we only keep samples with only one singlet bond crossing the boundary according to RSRG. result is shown in Fig. \ref{IPR_N_62_302_4_500}. For both strong and weak disorder, we find little dependence on size $N$.

\subsection{Entanglement energy}{\label{ee}}

\begin{figure}
\includegraphics[width=.5\textwidth]{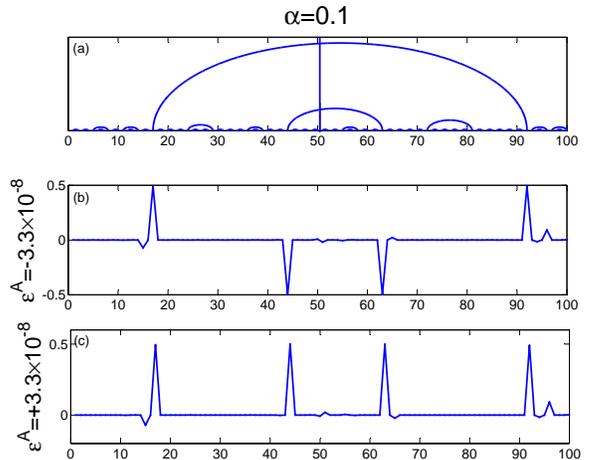}
\caption{\label{Klich_100_p1}Panel (a): Singlet bond configuration generated by RSRG method for an $N=100$ sample with $\alpha=0.1$ (strong disorder). Panels (b) and (c): Entangled modes whose entanglement energies are closest to zero. They are strongly localized at the entangled spins (consistent with RSRG prediction).}
\end{figure}

\begin{figure}
\includegraphics[width=.5\textwidth]{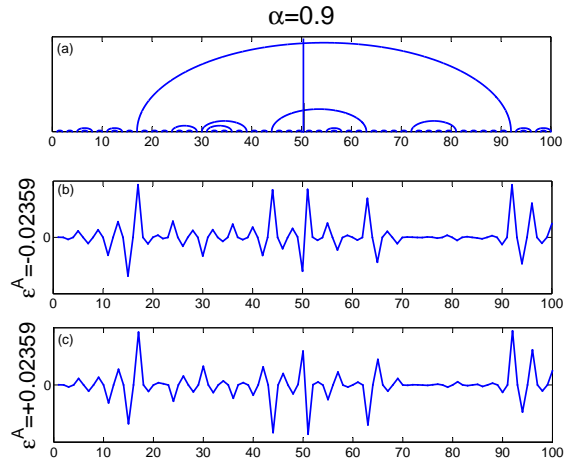}
\caption{\label{Klich_100_p9}Panel (a): Singlet bond configuration generated by RSRG method for an $N=100$ sample with $\alpha=0.9$ (weak disorder). Panels (b) and (c): Entangled modes whose entanglement energies are closest to zero. They are peaked at the entangled spins predicted by RSRG but much more spread out.}
\end{figure}

Now, we focus on entanglement energy. If we choose the number of spin sites, $N$, to be $2 \times $ even integer, then $N_A=N/2$ is even, RSRG would predict we will have an even number of entangled pairs of spins (possibly zero). In the following we focus on samples with exactly two of such entangled pairs based on RSRG, which then predicts there should be two modes whose entanglement energies are exactly zero, while all other entanglement energies should be $\pm\infty$.
We find that while never literally true, the above is close to being the case for strong disorder, but not quite so for weaker ones. As example we consider a system of $N=100, N_F= N_A=N_B=N/2=50$. Table \ref{t_p1_100} lists the entanglement energies of subsystem $A$, $\epsilon_k^A$,  and their corresponding probabilities, $n_k^A$, when we use $\alpha=0.1$; while Table \ref{t_p9_100} lists $\epsilon_k^A$ and $n_k^A$ when we use $\alpha=0.9$. We can see from comparison that for strong disorder, there is a pair of entanglement energies very close to zero, while (most) others have big magnitudes, consistent with RSRG prediction. For weaker disorder, on the other hand, the lowest entanglement energy is not as close to zero, and there are more entanglement energies that are of order 1. Fig. \ref{Klich_100_p1} (for strong disorder) and Fig. \ref{Klich_100_p9} (for weak disorder) shows RSRG generated singlet pair configuration for these two samples, and the two Klich eigenmodes with entanglement energy closest to zero.

\begin{table*}
\caption{\label{t_p1_100}Some of $\epsilon_k^A$'s and corresponding $n_k^A$'s for the case of  strong disorder ($\alpha=0.1$) and $N=100, N_F= N_A=N_B=N/2=50$. }
\begin{tabular}{c|cccccccccccccc}
$\epsilon_k^A$ &$- \infty$&$\cdots$& $-87.1$ & $-67.54$& $-28.84$ &$-3.0952$&$-3.3 \times 10^{-8}$&$+ 3.3 \times 10^{-8}$&$+3.0952$&$+28.84$&$+67.54$&$+87.1$&$\cdots$& $+ \infty$ \\
\hline
$n_k^A $ & $1$ & $\cdots$ &$\approx 1$& $ \approx 1$ &$\approx 1$& $0.9566$ & $0.5000$ & $ 0.5000$ & $0.0433$ &$2.9 \times 10^{-13}$& $\approx 0$ & $\approx 0$ &$\cdots $ & $0$ \\
\end{tabular}
\end{table*}

\begin{table*}
\caption{\label{t_p9_100}Some of $\epsilon_k^A$'s and corresponding $n_k^A$'s for the case of  weak disorder ($\alpha=0.9$) and $N=100, N_F= N_A=N_B=N/2=50$.}
\begin{tabular}{c|cccccccccccccc}
$\epsilon_k^A$ &$- \infty$&$\cdots$& $-25.79$ & $-16.47$& $-11.29$ &$-2.40$&$-0.02359$&$+0.02359$&$+2.40$&$+11.29$&$+16.47$&$+25.79$&$\cdots$& $+ \infty$\\
\hline
$n_k^A $ & $1$ & $\cdots$ &$\approx 1$& $ \approx 1$ & $0.9999$ & $0.9174$ & $0.5058$ & $ 0.4941$ & $0.0825$ &$1.2 \times 10^{-5}$& $7 \times 10^{-8}$ & $\approx 0$ &$\cdots $ & $0$ \\
\end{tabular}
\end{table*}

\begin{figure}
\includegraphics[width=.5\textwidth]{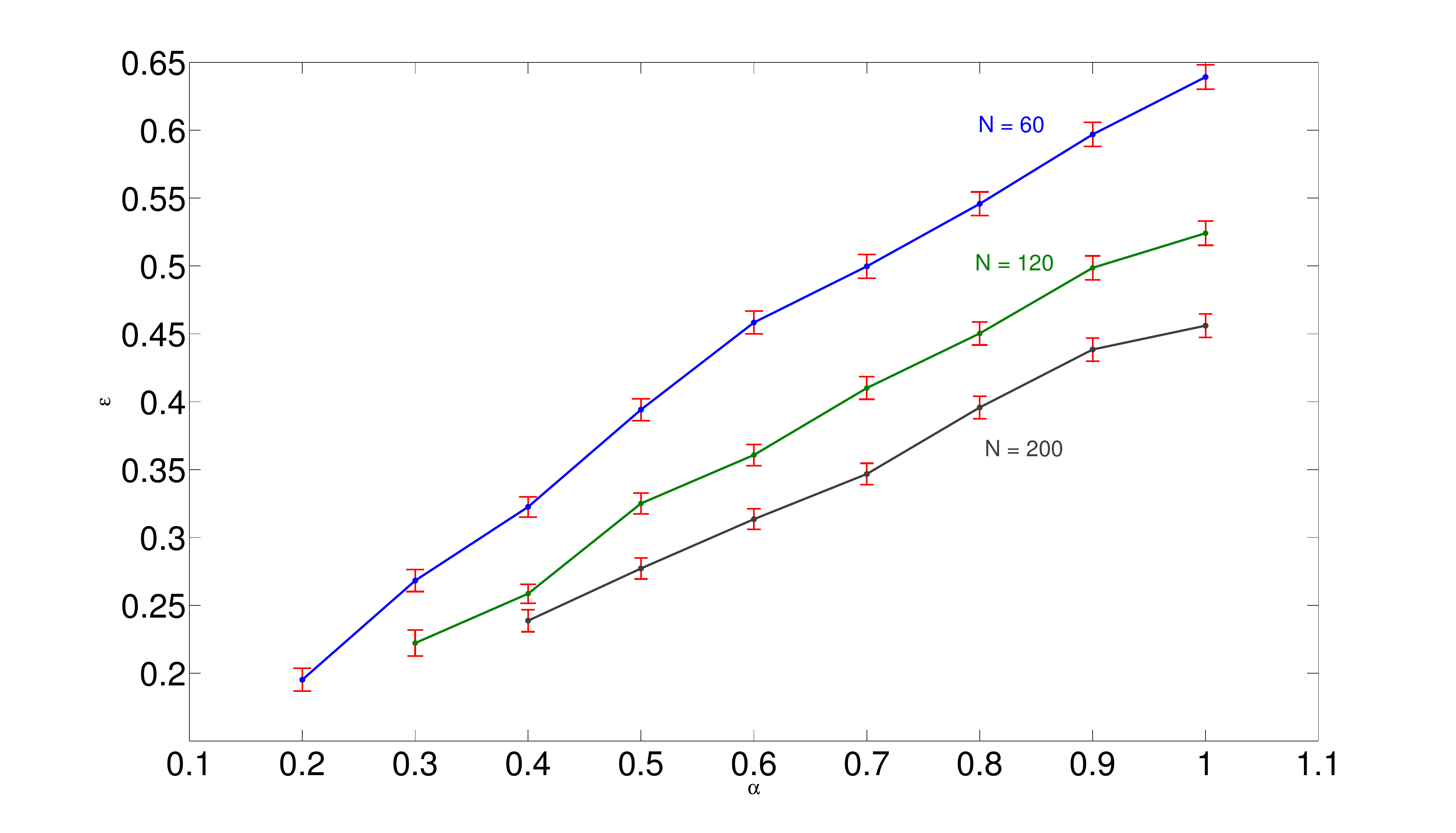}
\caption{\label{epsilon_alpha_2}[Color online] Average smallest (in magnitude) entanglement energy $\epsilon$ versus $\alpha$ for three samples of $N=60, N_F= N_A=N_B=N/2=30$ (blue), $N=120, N_F= N_A=N_B=N/2=60$ (green), and $N=200, N_F= N_A=N_B=N/2=100$ (gray). Average is calculated over $5000$ samples. Red line is  standard error. For strong disorder $\epsilon$ is close to zero and as we approach weak disorder regime it goes up.}
\end{figure}

To better quantify this point, we calculate smallest (in magnitude) entanglement energy averaged over many samples, for a range of $\alpha$'s. Fig. \ref{epsilon_alpha_2} shows the (averaged) smallest entanglement energy versus $\alpha$, for three samples of $N=60,120, 200$. As we can see in this figure, this energy is close to zero in strong disorder, while as we go to weaker disorder it increases, indicating the decreasing reliability of RSRG.

\begin{figure}
\includegraphics[width=.5\textwidth]{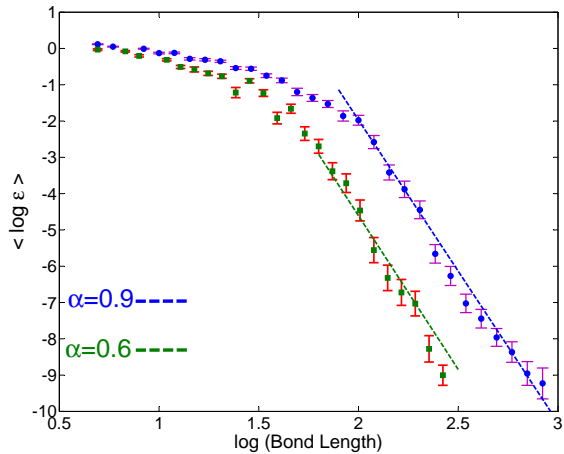}
\caption{\label{avlogepsilon_logbond}[Color online] Average logarithm of smallest entanglement energy ($\epsilon$) versus logarithm of bond length. For $\alpha=0.9$ we study samples ranging from size $N=100$ to $N=1000$ with interval of $\Delta N=100$. For $\alpha=0.6$, we study samples with $N=100, 200$ and $N=300$. For each $N$ we calculate smallest (in magnitude) $\epsilon$ and the corresponding bond length for $500$ samples. Logarithm of bond length is divided into $40$ segments, with $\log\epsilon$ within each segment averaged over. The blue line is best linear fit for $\alpha=0.9$ data with slop of $-8.3$ with standard deviation $0.5$. Green line is the best linear fit for the case of $\alpha=0.6$ data with slop of $-8.5$ with standard deviation $0.6$.}
\end{figure}

RSRG is expected to be asymptotically exact in the limit of long distance and low-energy, at least for thermodynamic properties.\cite{dsfisher} Here we would like to test if it is also the case for entanglement energy (the test for {\em real} excitation energies have been performed before\cite{Laflorencieetal,fishrtyoung}).
To this end we calculate the smallest entanglement energy versus bond length for a specific $\alpha$. Since we focus here on samples with two singlet bonds crossing the boundary, we identify the smallest (in magnitude) entanglement energy with the bond length of the longer of the two bonds. Since the distribution of the entanglement energies is very broad on logarithmic scale for large size systems, in Fig. \ref{avlogepsilon_logbond} we plot the average of $\log_{10}\epsilon$ vs. logarithm of bond length. The range of the latter is divided into $40$ segments, with $\log_{10}\epsilon$ within the same segment averaged over. We have also included data from samples with different systems sizes ($N$) in the averaging, as we find $\epsilon$ depends on bond length only. We find, qualitatively, the entanglement energy decreases with increasing bond length, and approaches zero as bond length $\rightarrow\infty$. This is consistent with asymptotical exactness of RSRG. More quantitatively,
we find that for two different disorder strengths, $\alpha=0.9$ and $\alpha=0.6$, $<\log_{10}\epsilon>$ depends on logarithm of bond length linearly (beyond certain length scale), indicating
\begin{equation}
\epsilon_{typical}\sim L^{-a},
\left\{
	\begin{array}{ll}
    	a = 8.3\pm 0.5, &  \alpha=0.9, \\
	a = 8.5\pm 0.6, &  \alpha=0.6,
	\end{array}
\right.
\end{equation}
where $L$ is bond length, $\epsilon_{typical}=e^{<\log\epsilon>}$ is the {\em typical} value of entanglement energy (in contrast to {\em average} value, which is often dominated by rare fluctuations for broad distributions).
Such power law behavior is consistent with the quantum critical nature of the random singlet phase. The corresponding exponent is the same for the two cases within error bars, indicating its universality. We note that within RSRG, the effective bond strength (and corresponding singlet to triplet excitation energy) scales with bond length in a (stretched) exponential fashion. Here we are studying {\em entanglement} energy, a {\em different} quantity, and find it scales with bond length in a power-law fashion. We would like to caution though that our study is limited to moderately large sizes, and cannot completely rule out the possibility that true asymptotic behavior could be different.

\section{Summary and Concluding Remarks}

In this paper we have studied the entanglement spectrum and in particular, the pairs of entangled modes of random spin-1/2 XX chains using a free fermion mapping, and compared them with predictions of real space renormalization group (RSRG) treatment. We find that RSRG results are qualitatively valid, but not quantitatively accurate, especially for modest disorder strength. On the other hand, in the large distance limit its prediction about entanglement energies becomes asymptotically exact. It would be interesting in the future to study if such asymptotic exactness still holds in the presence of relevant perturbations that drives the RSRG flow away from the random singlet fixed point, like dimerization.\cite{yang} On the other hand, the spreading of the entangled modes does not decrease with increasing system size; we interpret the profiles of these modes as the images of the {\em effective} spins that form singlets in the random singlet phase. This demonstrates the usefulness of studying the entangled modes, as such effective spins cannot be studies quantitatively within RSRG.

In a broader context, we argue that eigenstates of the reduced density matrices contain much useful information about entanglement, just like their spectra which are widely studied now. While the example presented here is a very simple one, we hope it serves as a starting point for future studies of entanglement taking advantage of eigenstates of the reduced density matrices.

\acknowledgements
This research is supported by the National Science Foundation through grant No. DMR-1004545.


\bibliography{apssamp}

\begin{thebibliography}{9}
\bibitem{pbcnote}
The periodic boundary condition (PBC) is often used in finite-size studies to maintain translation symmetry. Here translation symmetry is lost due to randomness, thus PBC has no advantage. In addition if we were to use PBC, dividing system into two would lead to two boundaries, complicating the analysis. The open boundary condition case has only one boundary, and is thus advantageous for entanglement studies.
\end{thebibliography}

\end{document}